\documentclass{article}\usepackage{alt-ciss2005}

\usepackage{amsmath,epsfig}

\begin{document}

\itwtitle{Impulse Radio Systems with Multiple Types of\\
Ultra-Wideband Pulses\small$\,^0$\\}

\itwauthor      {Sinan Gezici$^{1}$, Zafer Sahinoglu$^{2}$,
Hisashi Kobayashi$^{1}$, and H. Vincent Poor$^{1}$}
                {$^{1}$Department of Electrical Engineering, Princeton University, Princeton, NJ
08544, USA\\$^{2}$Mitsubishi Electric Research Laboratories, 201
Broadway, Cambridge, MA 02139, USA}

%  To add a second or third author, use the macros \itwsecondauthor and
%  \itwthirdauthor.
%
%  \itwsecondauthor{Giorgio Taricco}
%                 {Affiliation here}
%  \itwthirdauthor{Gavino Puddu}
%                 {Affiliation here}

\itwmaketitle

\footnotetext[0]{This research was supported by Mitsubishi
Electric Research Laboratories, Cambridge, MA, in part by the
National Science Foundation under grant ANI-03-38807, and in part
by the New Jersey Center for Wireless Telecommunications.}

\begin{itwabstract}
Spectral properties and performance of multi-pulse impulse radio
ultra-wideband systems with pulse-based polarity randomization are
analyzed. Instead of a single type of pulse transmitted in each
frame, multiple types of pulses are considered, which is shown to
reduce the effects of multiple-access interference. First, the
spectral properties of a multi-pulse impulse radio system is
investigated. It is shown that the power spectral density is the
average of spectral contents of different pulse shapes. Then,
approximate closed-form expressions for bit error probability of a
multi-pulse impulse radio system are derived for RAKE receivers in
asynchronous multiuser environments. The theoretical and
simulation results indicate that impulse radio systems that are
more robust against multiple-access interference than a
``classical" impulse radio system can be designed with multiple
types of ultra-wideband pulses.

\end{itwabstract}

\begin{itwpaper}

  \itwsection{Introduction}\label{sec:intro}

Impulse radio ultra-wideband (IR-UWB) systems hold great promise
for a variety of applications such as short-range high-speed data
transmission and precise location estimation. In such systems, a
train of pulses is transmitted and information is usually conveyed
by the position or the polarity of the pulses
\cite{scholtz}-\cite{Win2000}. In order to prevent catastrophic
collisions among different users and thus provide robustness
against multiple-access interference (MAI), each information
symbol is represented by a sequence of pulses; the positions of
the pulses within that sequence are determined by a pseudo-random
time-hopping (TH) sequence specific to each user \cite{scholtz}.

In ``classical" impulse radio, a single type of UWB pulse is
transmitted in all frames of all the users \cite{scholtz}. In
asynchronous multiuser environments, the autocorrelation function
of the pulse becomes an important factor in determining the
effects of the MAI \cite{sinan_WCNC04}. In order to reduce those
effects, UWB pulses with fast decaying autocorrelation functions
are desirable. However, such autocorrelation functions also result
in a considerable decrease in the desired signal part of the
receiver output in the presence of timing jitter
\cite{sinan_ICC04}. Moreover, when there is an exact overlap
between the pulses of two users, the MAI is usually very
significant. Hence, there is not much flexibility in choosing the
pulse shape in order to combat against interference effects.
However, in IR systems with multiple types of UWB pulses, MAI can
be mitigated by means of different types of UWB pulses with good
cross-correlation properties. Multi-pulse IR systems have recently
been proposed in \cite{kohno_UWBST04}. However, there has been no
theoretical analysis of those systems, in terms of their spectral
properties and bit error probability (BEP) performance, and no
quantitative investigation of the gains that can be obtained by
multiple types of UWB pulses.

In this paper, we consider an asynchronous multiuser environment
and analyze the BEP performance of a generic RAKE receiver over
frequency-selective channels. The results are valid for different
number of UWB pulse types, hence cover the single-pulse system as
a special case. We also analyze average power spectrum density
(PSD) of multi-pulse IR signals, and obtain a simple relation
between the Fourier transforms of the UWB pulses and the average
PSD of the transmitted signal.

\newcounter{sec}
\setcounter{sec}{2} The remainder of the paper is organized as
follows. Section \Roman{sec} describes the transmitted signal
model and Section \setcounter{sec}{3}\Roman{sec} analyzes its
spectral properties. In Section \setcounter{sec}{4}\Roman{sec},
the performance analysis of multi-pulse IR RAKE receivers is
presented. The simulation results are given in Section
\setcounter{sec}{5}\Roman{sec}, which is followed by the
concluding remarks in Section \setcounter{sec}{6}\Roman{sec}.

  \itwsection{Signal Model}

Consider a $K$-user environment with the following transmitted
signal from user $k$:
\begin{gather}\label{eq:tranSig}
s^{(k)}(t)=\frac{1}{\sqrt{N_f}}\sum_{j=-\infty}^{\infty}d^{(k)}_jb_{\lfloor
j/N_f\rfloor}^{(k)}p^{(k)}_j(t-jT_f-c_j^{(k)}T_c),
\end{gather}
where $p^{(k)}_j(t)$ is the UWB pulse transmitted in the $j$th
frame of user $k$, $b^{(k)}_{\lfloor j/N_f \rfloor}\in \{+1,-1\}$
is the equiprobable information bit, $T_f$ is the frame time,
$T_c$ is the chip interval, and $N_f$ is the number of
frames/pulses per information symbol. The time hopping (TH) code,
denoted by $c_j^{(k)}$, is modelled by a uniform distribution in
$\{0,1,...,N_c-1\}$, with $N_c=T_f/T_c$ being the number of chips
per frame, and $c_j^{(k)}$ and $c_i^{(l)}$ are independent for
$(j,k)\ne(i,l)$. Random polarity codes, $d^{(k)}_j$, are binary
random variables taking values $\pm1$ with equal probability, and
$d^{(k)}_j$ and $d^{(l)}_i$ are independent for $(j,k)\ne(i,l)$
\cite{eran1}. Use of random polarity codes helps reduce the
spectral lines in the power spectral density of the transmitted
signal \cite{paul} and mitigate the effects of MAI \cite{eran1}.
The receiver for user $k$ is assumed to know its polarity code.

Note the difference of the signal model in (\ref{eq:tranSig}) from
a classical IR system, in which the same pulse is used in all the
frames. In other words, the signal model in (\ref{eq:tranSig}) is
a more general formulation of an IR system, which reduces to the
original proposal in \cite{scholtz} when $p^{(k)}_j(t)=p(t)$
$\forall j,k$. We assume that there are $N_p$ different types of
pulses employed in the system and
$p^{(k)}_{j+iN_p}(t)=p^{(k)}_j(t)$ for any integer $i$. Also, for
simplicity of the expressions, we assume that $N_f$ is an integer
multiple of $N_p$.

  \itwsection{Power Spectrum Density Analysis}

In order to evaluate the spectral properties of the transmitted
signal, its (average) PSD needs to be calculated. Therefore, we
first calculate the autocorrelation function of $s(t)$ in
(\ref{eq:tranSig}) as follows$^3$\footnotetext[3]{We drop the user
index $k$ in this section, for notational convenience.}:
\begin{align}\label{eq:autoCorTXsig}\nonumber
& \phi_{ss}(t+\tau,t)=\textrm{E}\{s(t+\tau)s(t)\}=\\
&\frac{1}{N_f}\sum_{j=-\infty}^{\infty}
\textrm{E}\{p_{j}(t+\tau-c_jT_c-jT_f)p_{j}(t-c_jT_c-jT_f)\},
\end{align}
where we employ the fact that the random polarity codes are i.i.d.
random variables taking $\pm 1$ with equal probability.

From (\ref{eq:autoCorTXsig}), it is observed that $s(t)$ is not
wide-sense stationary (WSS) since the autocorrelation function is
not independent of $t$. However, note that $s(t)$ is a zero mean
cyclostationary process \cite{proakis} since $\phi_{ss}(t+\tau,t)$
is periodic with a period of $N_pT_f$. Therefore, we can obtain
the time-average autocorrelation function as
\begin{align}\label{eq:timeAvg_autoCor}\nonumber
\bar{\phi}_{ss}(\tau)&=\frac{1}{N_pT_f}\int_{0}^{N_pT_f}\phi_{ss}(t+\tau,t)dt\\
&=\frac{1}{N_pT_fN_f}\sum_{l=0}^{N_p-1}\int_{-\infty}^{\infty}p_l(t+\tau)p_l(t)dt,
\end{align}
the Fourier transform of which gives the average PSD as follows:
\begin{gather}\label{eq:PSD}
\Phi_{ss}(f)=\frac{1}{N_pT_s}\sum_{l=0}^{N_p-1}|P_l(f)|^2,
\end{gather}
where $T_s=N_fT_f$ is the symbol interval, and $P_l(f)$ is the
Fourier transform of $p_l(t)$.

Note from (\ref{eq:PSD}) that the average PSD of the signal is the
average value of the squares of the Fourier transforms of the
pulses. The dependence on the pulse spectra only is the result of
the pulse-based polarity randomization \cite{paul}, \cite{lehman}.
Moreover, note that there can be flexibility in shaping the PSD by
proper choice of UWB pulse types.

\itwsection{Performance Analysis}

Consider the following channel model for user $k$:
\begin{gather}\label{eq:chan_IR}
h^{(k)}(t)=\sum_{l=0}^{L-1}\alpha_l^{(k)}\delta(t-\tau_l^{(k)}),
\end{gather}
where $\alpha_l^{(k)}$ and $\tau_l^{(k)}$ are the fading
coefficient and the delay for the $l$th path of user $k$.

Using the channel model in (\ref{eq:chan_IR}) and the transmitted
signal in (\ref{eq:tranSig}), the received signal can be expressed
as
\begin{align}\label{eq:recSig}\nonumber
r(t)=\sum_{k=1}^{K}\frac{1}{\sqrt{N_f}}\sum_{j=-\infty}^{\infty}d^{(k)}_j&b_{\lfloor
j/N_f\rfloor}^{(k)}u_j^{(k)}(t-jT_f-c_j^{(k)}T_c\\
&\quad\quad-\tau_0^{(k)})+\sigma_nn(t),
\end{align}
with
\begin{gather}
u_j^{(k)}(t)\overset{\Delta}=\sum_{l=0}^{L-1}\alpha_l^{(k)}w^{(k)}_j(t-\tau_l^{(k)}+\tau_0^{(k)}),
\end{gather}
where $w^{(k)}_j(t)$ is the received UWB pulse in the $j$th frame
of user $k$, and $n(t)$ is a zero mean white Gaussian process with
unit spectral density.

We consider a generic RAKE receiver that can represent different
combining schemes, such as equal gain combining or maximal ratio
combining. It can be expressed as the correlation of the received
signal in (\ref{eq:recSig}) with the following template signal for
the $i$th information bit, where we consider the user $1$ as the
user of interest without loss of generality:
\begin{gather}\label{eq:tempSig}
s_i^{(1)}(t)=\sum_{j=iN_f}^{(i+1)N_f-1}d^{(1)}_jv^{(1)}_j(t-jT_f-c^{(1)}_jT_c),
\end{gather}
with
\begin{gather}
v^{(1)}_j(t)\overset{\Delta}=\sum_{l=0}^{L-1}\beta_lw^{(1)}_j(t-\tau_l^{(1)}),
\end{gather}
where $\beta_l$ denotes the RAKE combining coefficient for the
$l$th path. We assume $\tau_0^{(1)}=0$ without loss of generality.
Note that for a partial or selective RAKE receiver
\cite{cassiICC02}, the combining coefficients for those paths that
are not utilized are set to zero.

We assume that the TH codes are constrained to the set
$\{0,1,\dots,N_h-1\}$, where $N_hT_c+\tau^{(k)}_{L-1}<T_f$
$\forall k$, so that there is no inter-frame interference (IFI).
Then, using (\ref{eq:recSig}) and (\ref{eq:tempSig}), the decision
variable for detecting the $i$th bit of user $1$ can be obtained
as:
\begin{align}\label{eq:decVar}\nonumber
Y_i&=\int
r(t)s_i^{(1)}(t)dt\\
&=b_i^{(1)}\frac{1}{\sqrt{N_f}}\sum_{j=iN_f}^{(i+1)N_f-1}\phi_{u_j^{(1)}v_j^{(1)}}(0)+M_i+N_i,
\end{align}
with
\begin{gather}\label{eq:phi}
\phi_{u_i^{(k)}v_j^{(l)}}(x)\overset{\Delta}=\int
u_{i}^{(k)}(t-x)v_j^{(l)}(t)dt,
\end{gather}
where the first term in (\ref{eq:decVar}) is the desired signal
part of the output, $M_i$ is the MAI, and $N_i$ is the output
noise for the $i$th information bit. For simplicity of the
expressions, we drop the bit index $i$ and consider the 0th bit
without loss of generality, for the rest of the analysis.

\itwsubsection{Multiple-Access Interference}\label{sec:MAI}

Consider the MAI term $M$ in (\ref{eq:decVar}), which is the sum
of interference terms from $(K-1)$ users,
$M=\frac{1}{\sqrt{N_f}}\sum_{k=2}^{K}M^{(k)}$, where $M^{(k)}$ can
be expressed as
\begin{gather}\label{eq:M_i_k}
M^{(k)}=\sum_{j=0}^{N_f-1}\hat{M}_j^{(k)},
\end{gather}
with $\hat{M}_j^{(k)}$ denoting the MAI from user $k$ to the $j$th
frame of the first user. From (\ref{eq:recSig}),
(\ref{eq:tempSig}) and (\ref{eq:phi}), $\hat{M}_j^{(k)}$ can be
expressed as
\begin{align}\label{eq:M_hat_j_k}\nonumber
&\hat{M}_j^{(k)}=d_j^{(1)}\sum_{m=-\infty}^{\infty}d_m^{(k)}b_{\lfloor
m/N_f\rfloor}^{(k)}\\
&\phi_{u_m^{(k)}v_j^{(1)}}\left((m-j)T_f+(c_m^{(k)}-c_j^{(1)})T_c+\tau_0^{(k)}\right),
\end{align}
where $\tau_0^{(k)}$ denotes the amount of asynchronism between
user $k$ and the user of interest, user $1$, since we assume
$\tau_0^{(1)}=0$. In practical situations, the users in an IR-UWB
system are not synchronized. For example, in IEEE 802.15.3a
personal area networks, the MAI comes from neighboring
uncoordinated piconets. Therefore, we consider an asynchronous
scenario in which the amount of asynchronism is uniformly
distributed in a symbol interval; that is
$\tau_0^{(k)}\sim\mathcal{U}[0\,,\,N_fT_f)$ $\forall k$.

In order to obtain an approximate closed form expression for the
BEP, we assume large number of interferers with equal received
powers (perfect power control), and approximate the total MAI by a
zero mean Gaussian random variable using the standard Gaussian
approximation (SGA). Although this assumption may not be very
realistic in some situations, it will give us an idea about the
gains that can be obtained by multiple types of UWB pulses.
Moreover, if the number of users is not very large and the
received powers are unbalanced, we can employ the Gaussian
approximation in \cite{sinan_WCNC04} in order to obtain an
approximate BEP expression.

In order to calculate the variance of $M$ in (\ref{eq:decVar}), we
first consider that of $M^{(k)}$ in (\ref{eq:M_i_k}). Note that
$\textrm{E}\{\hat{M}_{j}^{(k)}\hat{M}_{l}^{(k)}\}=0$ for $j\ne l$
due to the random polarity codes. Therefore,
$\textrm{E}\{(M^{(k)})^2\}=\sum_{j=0}^{N_f-1}\textrm{E}\{(\hat{M}_j^{(k)})^2\}$.

From (\ref{eq:M_hat_j_k}), the variance of $\hat{M}_j^{(k)}$ can
be obtained as follows, after averaging over polarity
randomization and TH codes, and the delay of user $k$:
\begin{gather}
\textrm{E}\{(\hat{M}_j^{(k)})^2\}=\frac{\sigma_{M}^2(k,j)}{N_h^2},
\end{gather}
where
\begin{align}\label{eq:varMAIkj}\nonumber
\sigma_{M}^2(k,j)\overset{\Delta}=\frac{1}{T_fN_p}\sum_{m=j-N_p}^{j}\sum_{l=1-N_h}^{N_h-1}(N_h-|l|)\\
\int_{0}^{N_pT_f}\phi^2_{u_{m}^{(k)}v_{j}^{(1)}}\left((m-j)T_f+lT_c+\tau_0^{(k)}\right)d\tau_0^{(k)}.
\end{align}
Note that since there are $N_p$ different pulse shapes, it is
enough to integrate over $N_p$ frames, instead of the whole symbol
period.

For the classical IR system, where a single UWB pulse $w_0(t)$ is
employed, the result reduces, after some manipulation, to
\begin{align}\label{eq:var_M_hat_j_k_classical}\nonumber
\textrm{E}\{(\hat{M}_j^{(k)})^2\}=\frac{1}{T_fN_h^2}\sum_{l=1-N_h}^{N_h-1}(N_h-|l|)\\
\int_{-T_f}^{T_f}\phi^2_{u_0^{(k)}v_0^{(1)}}\left(lT_c+\tau_0^{(k)}\right)d\tau_0^{(k)}.
\end{align}

Note from (\ref{eq:varMAIkj}) and
(\ref{eq:var_M_hat_j_k_classical}) that the MAI term for the
classical IR system depends on the autocorrelation function of the
UWB pulse, whereas that for the multi-pulse IR system depends on
both the autocorrelation and the cross-correlation functions of
the pulses, which provides flexibility to design pulses with good
cross-correlation properties in order to reduce the effects of
MAI.

\itwsubsection{Output Noise}

The noise $N$ in (\ref{eq:decVar}) is distributed as
$\mathcal{N}\left(0\,,\,\sigma_n^2\int|s_{i}^{(1)}(t)|^2dt\right)$.
Using the expression in (\ref{eq:tempSig}) for $s_i^{(1)}(t)$, we
can obtain the distribution of $N$ for an IR system with $N_p$
different UWB pulses as follows:
\begin{gather}\label{eq:N_i_double}
N\sim\mathcal{N}\left(0\,,\,\sigma_n^2\frac{N_f}{N_p}\sum_{j=0}^{N_p-1}\phi_{v_j^{(1)}}(0)\right),
\end{gather}
where $\phi_{v_j^{(k)}}(x)\overset{\Delta}=\int
v_j^{(k)}(t-x)v_j^{(k)}(t)dt$ is the autocorrelation function of
$v_j^{(k)}(t)$.

\itwsubsection{Bit Error Probability}

Using the results in the previous sections, we obtain the
approximate BEP expression for an IR system employing $N_p$
different UWB pulses as follows:

\small
\begin{gather}\label{eq:BEP_Np} P_e\approx Q\left(\frac{
\frac{1}{\sqrt{N_p}}\sum_{j=0}^{N_p-1}\phi_{u^{(1)}_jv^{(1)}_j}(0)}
{\sqrt{\frac{1}{N_fN_h^2}\sum_{j=0}^{N_p-1}\sum_{k=2}^{K}\sigma^2_{M}(k,j)
+\sigma_n^2\sum_{j=0}^{N_p-1}\phi_{v_j^{(1)}}(0) }}\right)
\end{gather}\normalsize
for large $K$, where $\sigma^2_{M}(k,j)$ is as given in
(\ref{eq:varMAIkj}).

For the case of a single type of UWB pulse $w_0(t)$, the BEP can
be expressed as
\begin{gather}\label{eq:BEP_single}
P_e\approx Q\left(\frac{\phi_{u_0^{(1)}v_0^{(1)}}(0)}
{\sqrt{\frac{1}{N_fN_h^2}\sum_{k=2}^{K}\sigma_{M}^2(k)+\sigma_n^2\phi_{v_0^{(1)}}(0)}}\right),
\end{gather}
where \small\begin{gather}\label{eq:varM_asyc_single}\nonumber
\sigma_{M}^2(k)\overset{\Delta}=\frac{1}{T_f}\sum_{l=1-N_h}^{N_h-1}(N_h-|l|)
\int_{-T_f}^{T_f}\phi^2_{u_0^{(k)}v_0^{(1)}}\left(lT_c+\tau_0^{(k)}\right)d\tau_0^{(k)}.
\end{gather}\normalsize

From (\ref{eq:BEP_Np}) and (\ref{eq:BEP_single}), the improvements
in BEP as a result of the use of multiple UWB pulse types can be
quantified approximately.

\itwsection{Simulation Results}

\begin{figure}
\begin{center}
\includegraphics[width=0.5\textwidth]{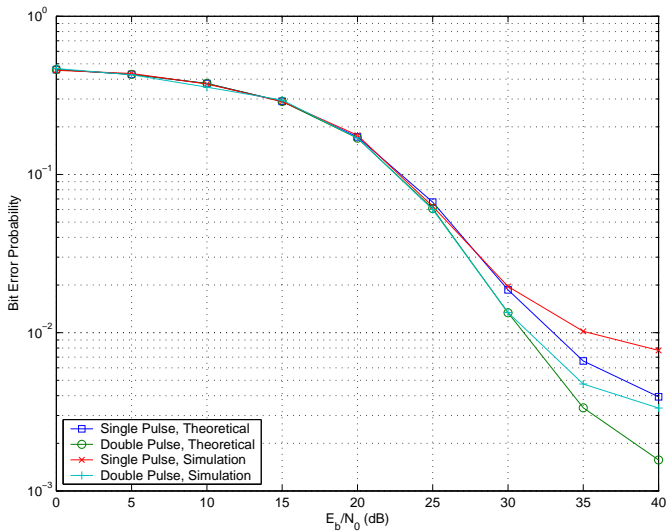}
\caption{BEP versus $E_b/N_0$ for single and double-pulse IR
systems.} \label{fig:BEP}
\end{center}
\end{figure}

In this section, we compare BEP performances of a single pulse and
a double-pulse IR system. In the double-pulse system, each user
transmits the $4$th and $5$th order modified Hermite pulses (MHPs)
\cite{kohno_UWBST04} alternately, whereas the single-pulse system
employs the $4$th order MHP in all the frames.

The systems parameters are $K=20$ users, $N_f=2$ frames per
symbol, $N_c=40$ chips per frame, $T_c=1$ ns, and $N_h=3$. We
consider a scenario, where the received energy of the interferers
is $5$ times larger than that of the user of interest. All the
channels have $L=20$ taps, which are generated independently
according to a channel model with exponentially decaying
($\textrm{E}\{|\alpha_l|^2\}=\Omega_0e^{-\lambda l}$) and
log-normally fading ($|\alpha_l|\sim\mathcal{LN}(\mu_l,\sigma^2)$)
channel amplitudes, random signs for channel taps, and exponential
distribution for the path arrivals with a mean $\hat{\mu}$. The
channel parameters are $\lambda=0.5$, $\sigma^2=1$, and
$\hat{\mu}=1.5$ ns, and $\mu_l$ can be calculated from
$\mu_l=0.5\left[\textrm{ln}(\frac{1-e^{-\lambda}}{1-e^{-\lambda
L}})-\lambda l-2\sigma^2\right]$, for $l=0,1,\ldots,L-1$.

Figure \ref{fig:BEP} shows the BEP performance of the all-RAKE
receivers \cite{cassiICC02} for the single and double-pulse
systems. Both the theoretical and the simulation results are
shown, which are in a quite good agreement, except for high SNR
values, where the SGA gives optimistic evaluation. From the plot,
the double-pulse system is observed to have a better performance
than the single-pulse system. From the expressions in Section IV,
the amount of MAI to the double-pulse system can be calculated to
be $\%20$ smaller than that of the single-pulse system. We can
obtain further gains by using more UWB pulse types and/or MHPs
that are several orders apart \cite{kohno_UWBST04}.

\itwsection{Conclusions}

We have obtained closed-form expressions for the average PSD and
approximate BEP of multi-pulse IR systems with pulse-based
polarity randomization. We have shown that by using different
types of UWB pulses, the effects of MAI can be mitigated, and we
have quantified the performance gains by using the approximate BEP
expressions.

The future work includes a more detailed performance analysis of
the multi-pulse IR system in the presence of IFI, considering the
effects of the MAI using more accurate approximations that do not
require large number of users or equal energy interferers.

\end{itwpaper}

\begin{itwreferences}

\bibitem{scholtz} M. Z. Win and R. A. Scholtz, ``Impulse radio:
How it works,'' \emph{IEEE Communications Letters,} 2(2): pp.
36-38, Feb. 1998.

\bibitem{scholtz4} M. Z. Win and R. A. Scholtz, ``On the energy capture of ultra-wide
bandwidth signals in dense multipath environments,'' \emph{IEEE
Communications Letters,} vol. 2, pp. 245-247, Sep. 1998.

\bibitem{welborn} M. L. Welborn, ``System considerations for
ultra-wideband wireless networks,'' \emph{Proc. IEEE Radio and
Wireless Conference,} pp. 5-8, Boston, MA, Aug. 2001.

\bibitem{scholtz1} R. A. Scholtz, ``Multiple access with time-hopping impulse modulation,''
\emph{Proc. IEEE Military Communications Conference (MILCOM
1993),} vol. 2, pp. 447-450, Bedford, MA, Oct. 1993.

\bibitem{FCC2002} U. S. Federal Communications Commission, ``First Report and Order
02-48,'' Washington, DC, 2002.

\bibitem{Win2000} M. Z. Win and R. A. Scholtz, ``Ultra-wide bandwidth
time-hopping spread-spectrum impulse radio for wireless
multiple-access communications,'' \emph{IEEE Transactions on
Communications,} vol. 48, issue 4, pp. 679-691, Apr. 2000.

\bibitem{sinan_WCNC04} S. Gezici, H. Kobayashi, H. V. Poor and A. F. Molisch,
``Performance evaluation of impulse radio UWB systems with
pulse-based polarity randomization in asynchronous multiuser
environments," \emph{Proc. IEEE Wireless Communications and
Networking Conference (WCNC 2004),} vol. 2, pp. 908-913, Atlanta,
GA, March 21-25, 2004.

\bibitem{sinan_ICC04} S. Gezici, H. Kobayashi, H. V. Poor and A. F. Molisch, ``The
trade-off between processing gains of an impulse radio system in
the presence of timing jitter,'' \emph{Proc. IEEE International
Conference on Communications (ICC 2004),} vol. 6, pp. 3596-3600,
Paris, France, June 20-24, 2004.

\bibitem{kohno_UWBST04} H. Harada, K. Ikemoto and R.
Kohno, ``Modulation and hopping using modified Hermite pulses for
UWB communications," \emph{Proc. IEEE Conference of Ultra Wideband
Systems and Technologies (UWBST 2004),} pp. 336-340, Kyoto, Japan,
May 2004.

\bibitem{eran1} E. Fishler and H. V. Poor, ``On the tradeoff between two types of processing
gain," \emph{Proc. 40th Annual Allerton Conference on
Communication, Control, and Computing,} Monticello, IL, Oct. 2-4,
2002.

\bibitem{paul} Y.-P. Nakache and A. F. Molisch, ``Spectral shape of UWB
signals influence of modulation format, multiple access scheme and
pulse shape," \emph{Proc. IEEE Vehicular Technology Conference,
(VTC 2003-Spring),} vol. 4, pp. 2510-2514, Jeju, Korea, April
2003.

\bibitem{lehman} N. H. Lehmann and A. M. Haimovich, ``The power spectral
density of a time hopping UWB signal: A survey,'' \emph{Proc. IEEE
Conference on Ultra Wideband Systems and Technologies (UWBST
2003)}, pp. 234-239, Reston, VA, Nov. 2003.

\bibitem{proakis} J. G. Proakis, \emph{Digital Communications,}
Mc Graw Hill, 4th edition, 2000.

\bibitem{cassiICC02} D. Cassioli, M. Z. Win, F. Vatalaro and A. F.
Molisch, ``Performance of low-complexity RAKE reception in a
realistic UWB channel," \emph{Proc. IEEE International Conference
on Communications (ICC 2002),} vol. 2, pp. 763-767, New York, NY,
April 28-May 2, 2002.

\end{itwreferences}

\end{document}